# TOWARDS DIRECT NUMERICAL SIMULATION OF A 5X5 ROD BUNDLE

**Adam Kraus\* and Elia Merzari**
Penn State University
State College, PA 16801
ark221@psu.edu; ebm5351@psu.edu

**Thomas Norddine**
École des Ponts ParisTech
6-8 avenue Blaise-Pascal, Cité Descartes, 77455, Champs-sur-Marne, Marne-la-Vallée cedex 2
thomas.norddine@eleves.enpc.fr

**Oana Marin**
Argonne National Laboratory
9700 S. Cass Avenue, Lemont IL 60439, USA
oanam@mcs.anl.gov

**Sofiane Benhamadouche**
Electricite de France
78401 Chatou Cedex, France
sofiane.benhamadouche@edf.fr

**ABSTRACT**

Rod bundle flows are prevalent in nuclear engineering for both light water reactors (LWR) and advanced reactor concepts. Unlike canonical channel flow, the flow in rod bundles presents some unique characteristics, notably due to the inhomogeneous cross section which can present different local conditions of turbulence as well as localized effects characteristic of external flows. Despite the ubiquity of rod-bundle flows and the decades of experimental and numerical knowledge acquired in this field, there are no publicly available direct numerical simulations (DNS) of the flow in multiple-pin rod bundles with heat transfer. A multiple-pin DNS study is of great value as it would allow for assessment of the reliability of various turbulence models in the presence of heat transfer, as well as allow for a deeper understanding of the flow physics. We present work towards DNS of the flow in a square 5x5 rod bundle representative of LWR fuel. We consider standard configurations as well as configurations where the central pin is replaced with a guide thimble. We perform simulations in STAR-CCM+ to design the numerical DNS, which is to be conducted using the open source spectral element code Nek5000. Large Eddy Simulations are also performed in Nek5000 to confirm that the resolution requirements are adequate. We compare results from STAR-CCM+ and Nek5000, which show very good agreement in the wide gaps with larger discrepancies in the narrow gaps. In particular, evidence of a gap vortex street is seen in the edge subchannels in LES but is not predicted by STAR-CCM+.

---

\* Also with Argonne National Laboratory

## 1. INTRODUCTION

Rod bundles are an essential component of most nuclear reactor concepts. The external flow over arrays of cylindrical rods is therefore of great interest to nuclear engineers and reactor designers. The prediction of the temperature distribution within the rod bundle is of particular importance. Unlike in canonical channel flows, the flow in rod bundles presents some unique characteristics that have been evaluated over decades of experimental and numerical research. Among the most interesting characteristics of rod bundles versus canonical flows is certainly the inhomogeneous cross section which can present different local conditions of turbulence as well as the presence of localized effects characteristics of external flows [1]. Historically, system codes and subchannel codes have been extensively used in the thermal-hydraulic design and analysis of reactor fuel assemblies, but they are limited by the need of empirical correlations derived from representative experimental results over a range of parameters. Computational Fluid Dynamics (CFD) has emerged in the last three decades as a promising solution to predict flow and heat transfer in rod bundles. Industry relies routinely on Reynolds Averaged Navier-Stokes (RANS) approaches to predict the flow in rod bundles with Light Water Reactor (LWR) spacer grids [2]. The simulation of wire-wrapped rod bundles has also been an active topic of research. RANS retains some limitations due to the need to provide closure relationships and benefits from comparison with turbulence resolving techniques. Such high fidelity techniques, including direct numerical simulations (DNS) or highly wall-resolved large eddy simulations (LES), allow for an accurate computation of the flow and heat transfer phenomena in nuclear assemblies, without the need for empirical correlations. Their cost, however, scales with the Reynolds number and remains prohibitive for realistic geometries.

We note that despite the wealth of literature available on rod bundles, there are no publicly available DNS of the flow in multiple-pin rod bundles with heat transfer. In fact, there are only few data available for single-pin infinite-lattice bundles, which are of limited use in practical applications as wall effects in the cross section are always important and instantaneous transverse periodic boundary conditions are not realistic. A multiple-pin DNS study is of great value to the engineering community as it would allow for the assessment of the reliability of various turbulence models in the presence of heat transfer. Moreover, such simulation would provide an invaluable source of data for a deeper understanding of the flow physics. In this paper we present work toward the direct numerical simulation of the flow in a 5 by 5 rod bundle representative of LWR fuel. Similar geometries have been used in a number of computational and experimental benchmarks [1,2,3]. We consider standard configurations as well as configurations where the central pin is removed and replaced with a guide thimble. The thimble is larger in diameter than the pin. It adds significantly to the non-uniformity of the flow in the cross section and changes the local gap Reynolds number. We perform simulations in STAR-CCM+ [4] to design the numerical DNS experiment to be conducted using the open source spectral element code Nek5000 [5]. Large Eddy Simulations are also performed in Nek5000 to confirm that the resolution requirements are adequate. We also analyze the anisotropy of the Reynolds stress tensor and compare results from STAR-CCM+ and Nek5000. The numerical methods are summarized in Section 2. The results from STAR-CCM+ used for the design of the numerical experiment are summarized in Section 3 while numerical results for Nek5000 are summarized and compared with those from STAR-CCM+ in Section 4. Finally, a brief discussion is provided in Section 5.

## 2. METHODS

Let us consider the velocity (Eq. (1)), continuity (Eq. (2)) and energy (Eq. (3)) equations that describe incompressible flow of a Newtonian fluid in the absence of other body or external forces.

$$\frac{\partial u_i}{\partial t} + \frac{\partial}{\partial x_j}(u_i u_j) = -\frac{1}{\rho}\frac{\partial p}{\partial x_i} + \frac{\partial}{\partial x_j}\left[\nu\left(\frac{\partial u_i}{\partial x_j} + \frac{\partial u_j}{\partial x_i}\right)\right] \qquad (1)$$

$$\frac{\partial u_i}{\partial x_i} = 0 \tag{2}$$

$$\rho c_p \left(\frac{\partial T}{\partial t} + u_j \frac{\partial T}{\partial x_j}\right) = \frac{\partial}{\partial x_j}\left(\lambda \frac{\partial T}{\partial x_j}\right) \tag{3}$$

where ρ is the constant density of the fluid, ν represents the kinematic viscosity, λ is the thermal conductivity and $c_p$ the heat capacity. Implicit summation applies. Let us also note here, that if the fluid viscosity does not depend on temperature (a valid assumption for the cases that are considered for the proposed methodology), velocity equations are completely uncoupled to the temperature field. For the sake of simplicity, thermal conductivity and the heat capacity will also be considered constant. Using the Reynolds averaging operator, one may partition instantaneous fields into an average part (time average for statistically steady flows) and a fluctuating part (i.e. $u_i = \bar{u}_i + u'_i$). When this is applied to the Navier-Stokes equations, the RANS equations are obtained. The additional terms that appear in velocity and energy equations from this averaging represent respectively the Reynolds stress tensor ($-\overline{u'_i u'_j}$) and the turbulent heat flux vector ($-\overline{u'_j T'}$). Different closures for those terms lead to the variety of RANS and turbulent heat flux models available in literature.

## 2.1. Advanced RANS computation STAR-CCM+

Pseudo-2D calculations [6] have been performed with the commercial finite volume code STAR-CCM+ [1] to estimate the Taylor microscale ($\lambda = \sqrt{\frac{15k\nu}{\varepsilon}}$) and the Kolmogorov scale ($\eta = \left(\frac{\nu^3}{\varepsilon}\right)^{\frac{1}{4}}$). The results have been used to design the LES and DNS numerical experiments performed with Nek5000 [7]. The DNS mesh has been designed such that the distance between Gauss-Lobatto-Legendre (GLL) points is smaller than $\min(\frac{\lambda}{2}, \eta)$.

An Elliptic Blending Reynolds Stress Model (EBRSM) [8, 9] has been used, and the following set of coupled equations has been solved:

$$\frac{\partial R_{ij}}{\partial t} + u_k \frac{\partial R_{ij}}{\partial x_k} = -(R_{ij}S_{ji} + R_{ij}S_{ij}) - \frac{\partial}{\partial x_k}\left(\left(C_s \frac{k}{\varepsilon} R_{km} + \nu\right)\frac{\partial R_{ij}}{\partial x_m}\right) + (1-\alpha^3)\left(\Phi^w_{ij} - \varepsilon^w_{ij}\right) + \alpha^3\left(\Phi^h_{ij} - \varepsilon^h_{ij}\right) \tag{4}$$

The main idea of the EBRSM, based on Durbin's elliptic relaxation approach [10], is to correctly predict the wall asymptotic behavior for the pressure strain term Φ and to represent the blocking of the wall normal fluctuations. The outer wall contribution is based on the "SSG" model of Speziale, Sarkar, and Gatski [11].

$$\varepsilon^w_{ij} = \frac{\bar{R}_{ij}}{k}\varepsilon \quad \text{and} \quad \varepsilon^h_{ij} = \frac{2}{3}\varepsilon\delta_{ij} \tag{5}$$

$$\Phi^w_{ij} = -5.0\frac{\varepsilon}{k}\left(R_{ik}n_j n_k + R_{jk}n_i n_k - \frac{1}{2}\left(R_{kl}n_k n_l n_i n_j - R_{kl}n_k n_l \delta_{ij}\right)\right) \tag{6}$$

$$\Phi^h_{ij} = -C_1\rho\varepsilon a_{ij} - C_{1s}P_k a_{ij} + \left(C_3 - C_{3s}\sqrt{a_{kl}a_{kl}}\right)\rho k S_{ij} + C_4\rho k\left(a_{ik}S_{jk} + a_{jk}S_{ik} - \frac{2}{3}a_{kl}S_{kl}\delta_{ij}\right) + C_5\rho k\left(a_{ik}W_{jk} + a_{jk}W_{ik}\right) \tag{7}$$

The elliptic blending factor, α, is the solution of the following elliptic equation:

$$\alpha - L^2 \nabla^2 \alpha = \frac{1}{k} \tag{8}$$

with $n_i = \frac{(\nabla \alpha)_i}{\|(\nabla \alpha)\|}$, the unit normal of the α gradient, and $L = C_L max\left(\frac{k^{3/2}}{\varepsilon}, C_\eta \left(\frac{\nu^3}{\varepsilon}\right)^{\frac{1}{4}}\right)$ with $C_L = 0.133$ and $C_\eta = 80$.

Since the EBRSM is a near wall model, $y^+ = 0.5$ and $x^+ = 3.5$ (i.e. the transverse direction) have consequently been targeted and checked *a posteriori*. A 2D mesh composed of quadrilateral elements was created with STAR-CCM+ and then extruded to form hexahedra, as seen in Figure 1. This resolution was confirmed to be adequate through mesh convergence studies using two other meshes. A periodic boundary condition has been set in the stream-wise direction and a constant bulk velocity is imposed during the simulation, which is steady. A second-order segregated space-centered scheme was used. Physical quantities were averaged after convergence to get rid of potential numerical fluctuations, although residuals of all quantities were converged below $10^{-4}$.

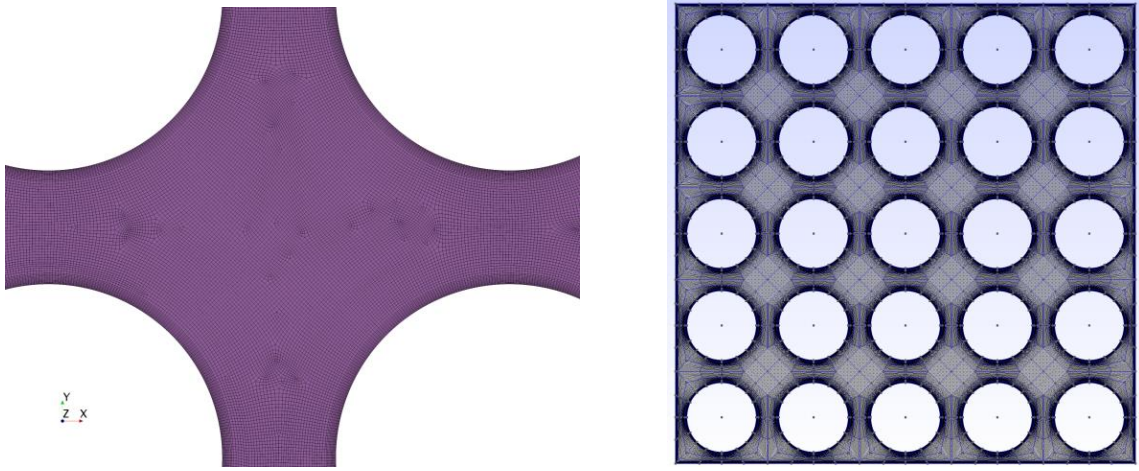

**Figure 1. Pseudo-2D periodic mesh used with STAR-CCM+ (left) and mesh element structure from Gmsh used in Nek5000 runs (right).**

## 2.2. LES with Nek5000

The LES simulations have been performed with the high-order spectral element code Nek5000 [5], and the planned DNS calculations will also be performed with Nek5000. The Spectral Element Method (SEM), introduced by Patera [12] is a subclass of the Galerkin methods. SEM combines the geometric flexibility of finite elements with the minimal numerical dispersion and dissipation of spectral methods, making it well suited for DNS and highly resolved LES computations of simple to complex geometries. The domain is discretized into E curvilinear hexahedral elements, in which the solution is represented as a tensor-product of Nth-order Lagrange polynomials based on the GLL nodal points, leading to a total of $n \approx E(N + 1)^3$ degrees or freedom per scalar field. Nek5000 was designed from the outset for distributed-memory platforms. It is highly parallel and has been recently applied to a variety of problems to gain unprecedented insight into the physics of turbulence in complex flows [13]. Time-stepping is semi-implicit: the viscous terms of the NS equations are treated implicitly (for the current simulations a second-order backward differentiation has been used, BDF2), while non-linear terms are treated by a third order extrapolation (EXT3) scheme. The main bottleneck of these computations is the resolution of the elliptic problem for the governing pressure. The discrete Poisson equation is solved using a variational multigrid GMRES method with *local* overlapping Schwarz methods for element-based smoothing at resolution N and ≈ N/2, coupled with a *global* coarse-grid problem based on linear elements. The LES formulation relies on explicit filtering of modes k = N - 1 and k = N, which provides an energy drain at the unresolved grid scale, similar to deconvolution LES models [14]. For the present LES simulations, 1% filtering of the last mode has been used. For well-resolved regions, the action of the filter is void and spectral accuracy is retained.

## 3. DESIGN OF THE NUMERICAL EXPERIMENT

### 3.1. Computational Domain and Physical Conditions

The geometry is a 5x5 square array of cylinders surrounded by a square wall. Dimensions are normalized such that the pin diameter is unity. Pitch is 1.326D, which is a typical pitch used in Pressurized Water Reactors (PWRs), and the domain width is 6.632D. The thimble diameter is roughly 1.278D. The domain length for the LES cases is 3.158D, which was chosen based on some sensitivity studies and review [1]. The problem was run non-dimensionally such that the density and bulk velocity were both unity. Thus the viscosity is the only driving parameter for the flow conditions and the pin-based Reynolds number is $1/\mu$. The problem was run at a Reynolds number of 19 000 based on the rod diameter.

For LES, a CFL number of 0.5 was targeted with an adaptive timestepper before beginning averaging, i.e. before achieving fully developed turbulence in the domain. Then a constant time step ensuring mean CFL ~ 0.5 was used for time averaging. In order to hasten the process of averaging the physical quantities, the turbulent quantities were spatially averaged over the stream-wise direction, which is a direction of homogeneity. The domain is invariant by $\pi/2$ rotations and the resulting domains have one symmetry plane. Thus one can average over 1/8 of the spatial domain.

### 3.2. Mesh Characteristics

The LES mesh consists of 1 038 880 curvilinear hexahedral elements and the polynomial order was set to 8, thus 531 906 560 gridpoints have been used for the simulation. The mesh was designed with a target dimensionless wall distance of $y^+ = 0.5$. The orthoradial ($x^+$) and stream-wise ($z^+$) dimensionless wall distance were set to 3.5 and 10 respectively. The open source mesh generator Gmsh was used to generate the hexahedral elements, as seen in Figure 1, and the GLL points were generated internally with Nek5000. Figure 2 shows that the mesh sizing used in the LES study is generally below the Taylor microscale (as estimated with STAR-CCM+) in the majority of the flow and so the LES can be considered well resolved. The Kolmogorov scales, which were also extracted from the STAR-CCM+ runs, can provide an estimate as to how much further refinement is necessary for DNS.

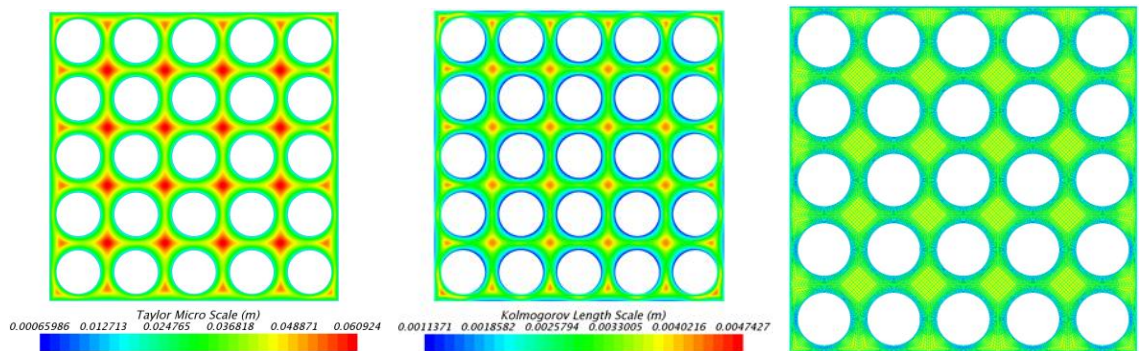

**Figure 2. Taylor length scale (left), Kolmogorov length scale (center), and LES mesh size compared to Kolmogorov scale (right, color scale is same as center).**

## 4. RESULTS

### 4.1. Standard 5x5 Configuration

The basic features of the flow field for the standard configuration are shown in Figure 3. Generally there is only little difference between the interior subchannels, while the edge and corner subchannels feature reduced flow and more complex behavior. Figure 4 demonstrates the profiles for the normal and shear Reynolds stress components for the EBRSM. These demonstrate the general wall peaking of the stress components, of which w'w' is largest, which is a feature of wall dominated flows.

To provide more direct comparisons, various quantities were extracted at six averaging lines, as shown in Figure 3. The first three lines proceed radially outward from the center pin along the x-axis, normal to the domain outer wall, i.e. in the "narrow gaps." Line one is next to the center and line three is in the edge subchannel. The last three lines proceed similarly along the diagonal, i.e. in the "wide gaps," with line six in the corner subchannel. In all of the plots, the coordinates begin at the wall closest to the domain center and increase radially.

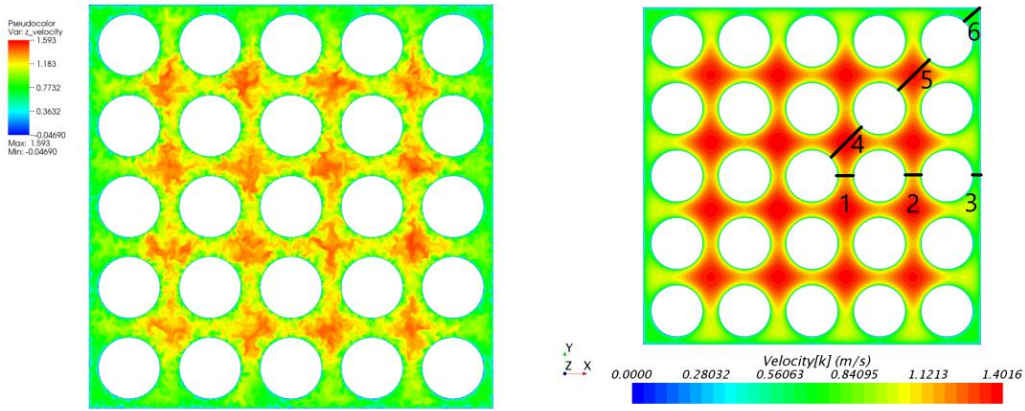

**Figure 3. Instantaneous axial velocity from LES (left) and steady-state axial velocity from EBRSM (right). Right plot demonstrates the lines used for averaging and plotting.**

Figure 5 provides the non-dimensional velocities in each of the six averaging lines. Only the innermost half-channel widths are shown. The laminar sublayer and log layer are clearly demonstrated. The "law of the wall" is matched very well by both LES and EBRSM approaches in the wide gap channels. This clearly shows that the EBRSM has been built for wall dominated flows and reproduces accurately the wall-turbulence for fully developed flows. Some deviations are however seen in the narrow gaps but they are relatively small. In both the corner and edge subchannels, the local Reynolds numbers are too small relative to the shear for there to be a substantial log layer, and large deviations are seen. These suggest that adaptive wall functions or "low-y+" approaches may be important for RANS turbulence models to achieve acceptable predictions in these regions.

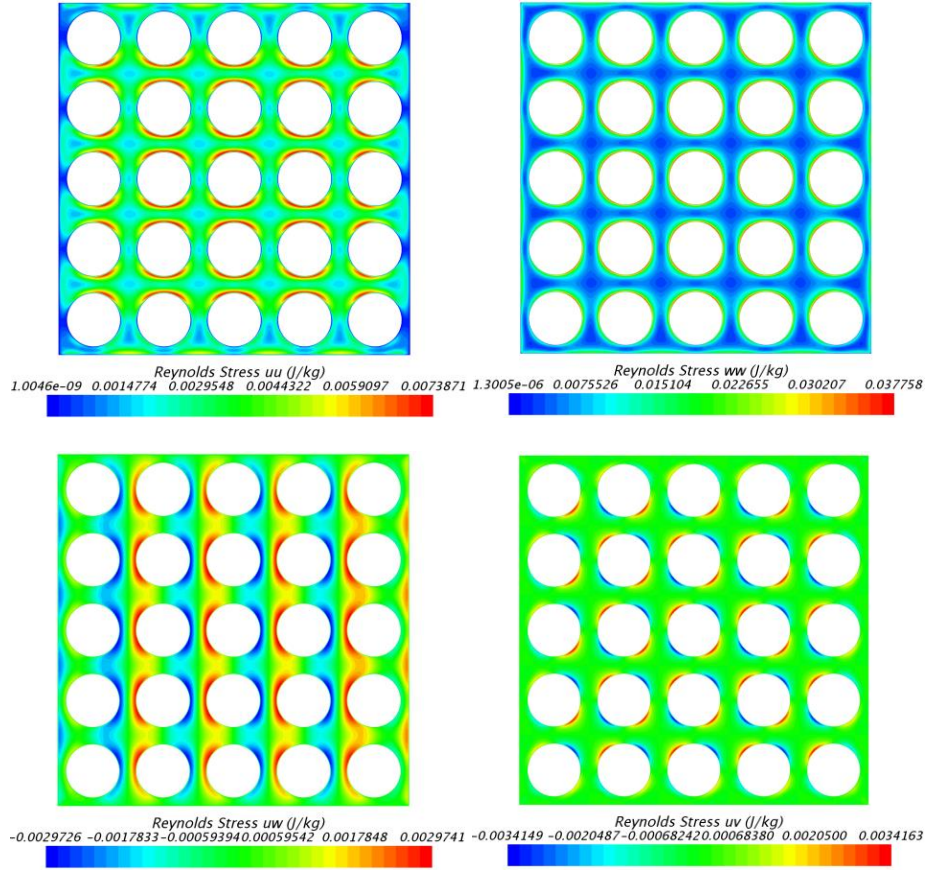

**Figure 4. Normal (top) and shear (bottom) Reynolds stress components from EBRSM.**

The normal Reynolds Stresses along the averaging lines (Figure 6) are roughly symmetric between the channel walls in the interior subchannels, whereas a clear tail-off is seen in the corner subchannel. All Reynolds stresses have been normalized to the peak LES stress, which occurs in the stream-wise direction. The wide gap predictions of EBRSM are in very good agreement with LES. It is readily seen though that the EBRSM model better matches the LES results when the channels are wider and the local Reynolds number is higher. Performance is better for the wide gaps than the narrow gaps, and deteriorates moving outward from the center. The shear Reynolds stresses are provided in Figure 7. The EBRSM model again yields excellent agreement in the wide gaps, with worse agreement in the narrow gaps. This is clearly shown along line 3. In general, the profile shapes match well.

The behavior in the edge subchannel, i.e. line 3, merits further consideration. In this region, large discrepancies are seen from EBRSM as compared to LES. The normal stress is much larger in LES and does not display the dual-peak form of most of the other components but rather a central peak. Some of this may stem from the known complexity of flow in narrow-gap rod banks due to the gap vortex street [15]. Figure 8 shows a snapshot of the flow in the edge channel centerline (plane perpendicular to line 3). We observe a sinusoidal path for the stream-wise velocity typical of the gap vortex flow. The peaking of the normal azimuthal stresses in the center of the domain is also consistent. As indicated by Merzari et al. [13], the gap vortex street can occur on the edge channels while being absent (or not dominant) in the center channels. This seems to occur here, we however note that to accurately predict the flow in the edge channels we will likely need a longer computational domain: in fact the current length is sufficient only to predict a single flow structure in the axial direction. Future DNS calculations will be performed with a longer computational domain to better represent the structures or with a modified edge channel dimension to avoid the presence of the gap vortex street. A DNS in this geometry should prove useful in elucidating the physical mechanism behind this complex behavior.

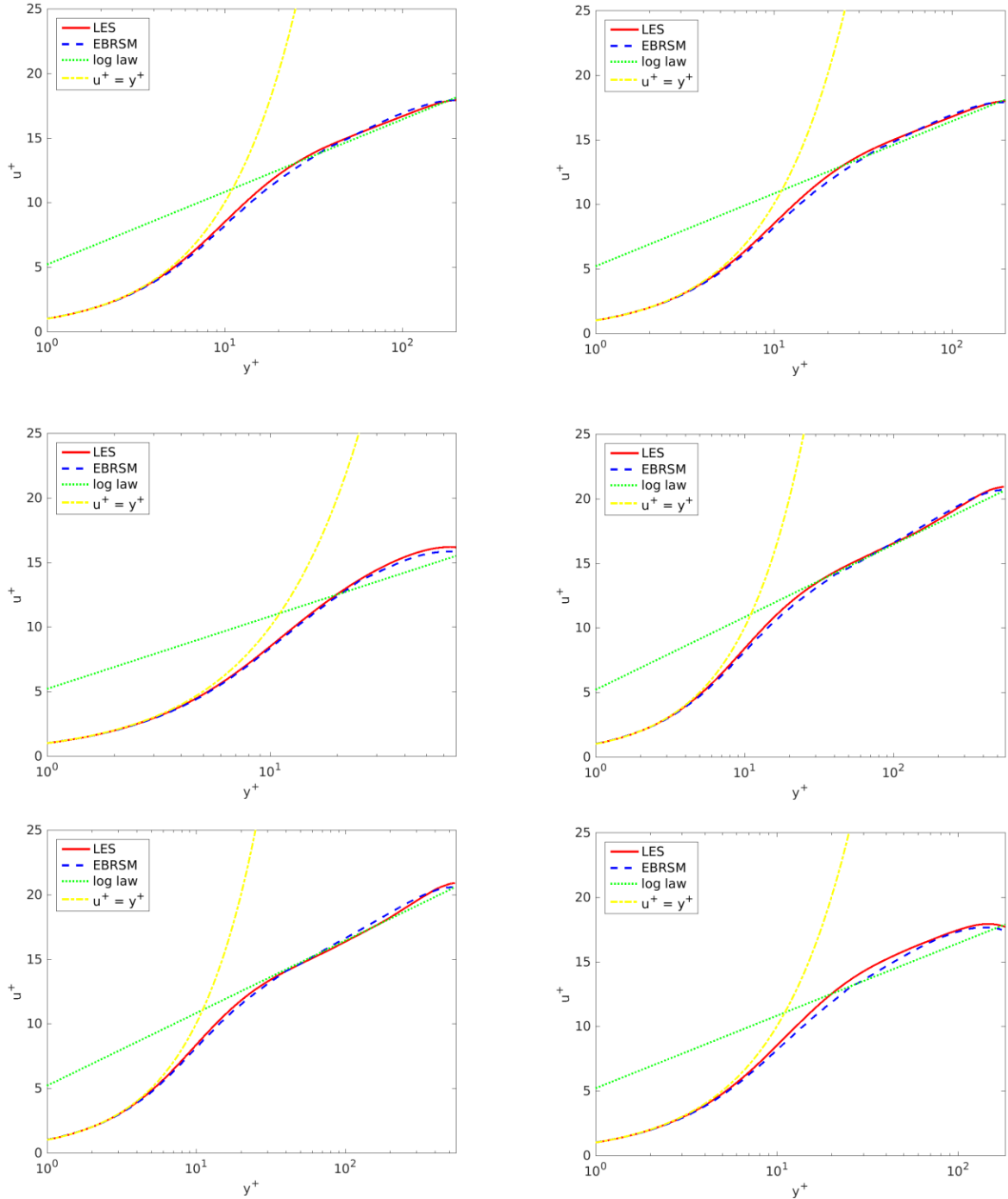

**Figure 5. Non-dimensional velocity profiles across the channel widths for lines 1 (top left) through 6 (bottom right).**

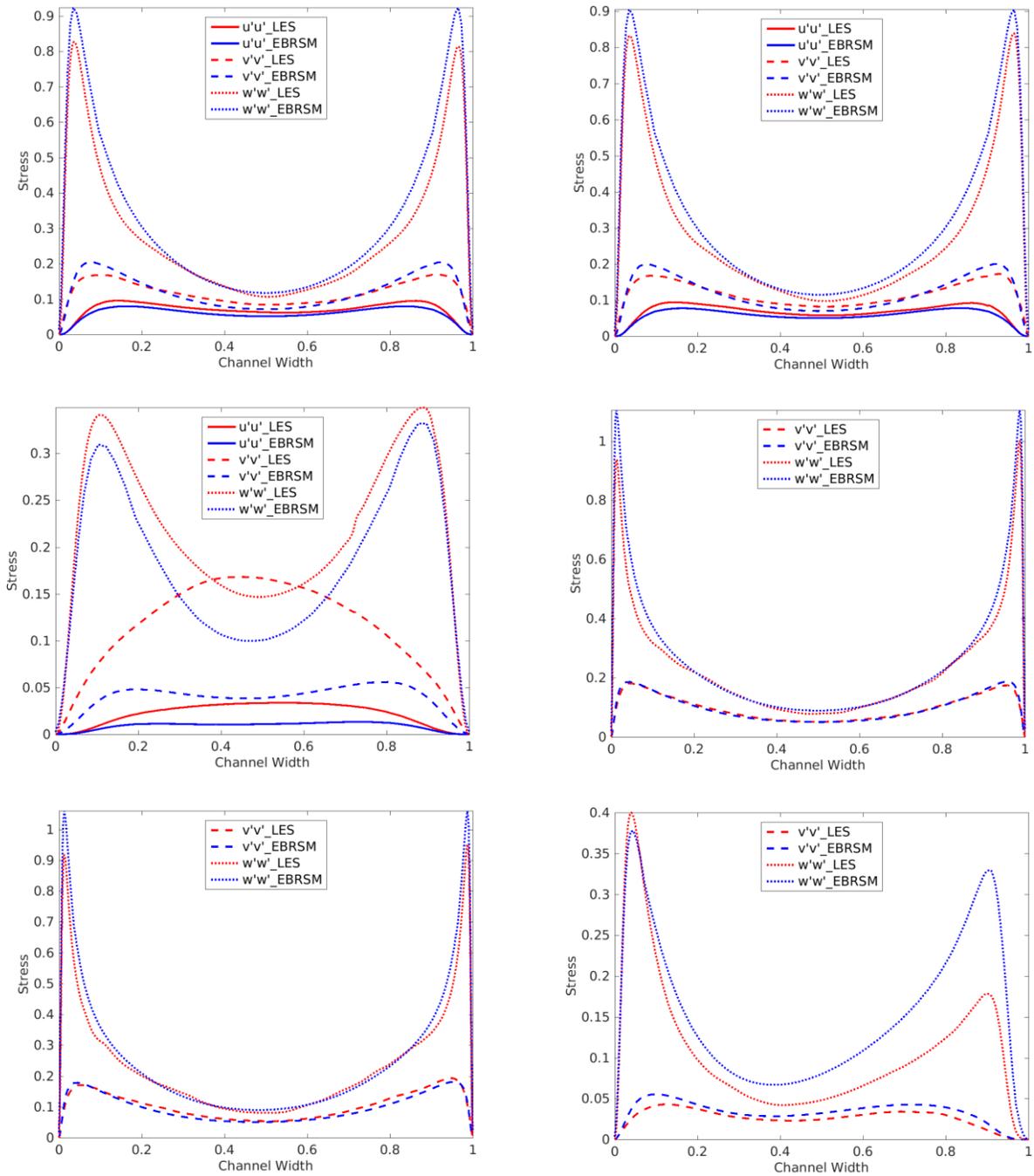

**Figure 6. Reynolds normal stresses for lines 1 (top left) through 6 (bottom right).**

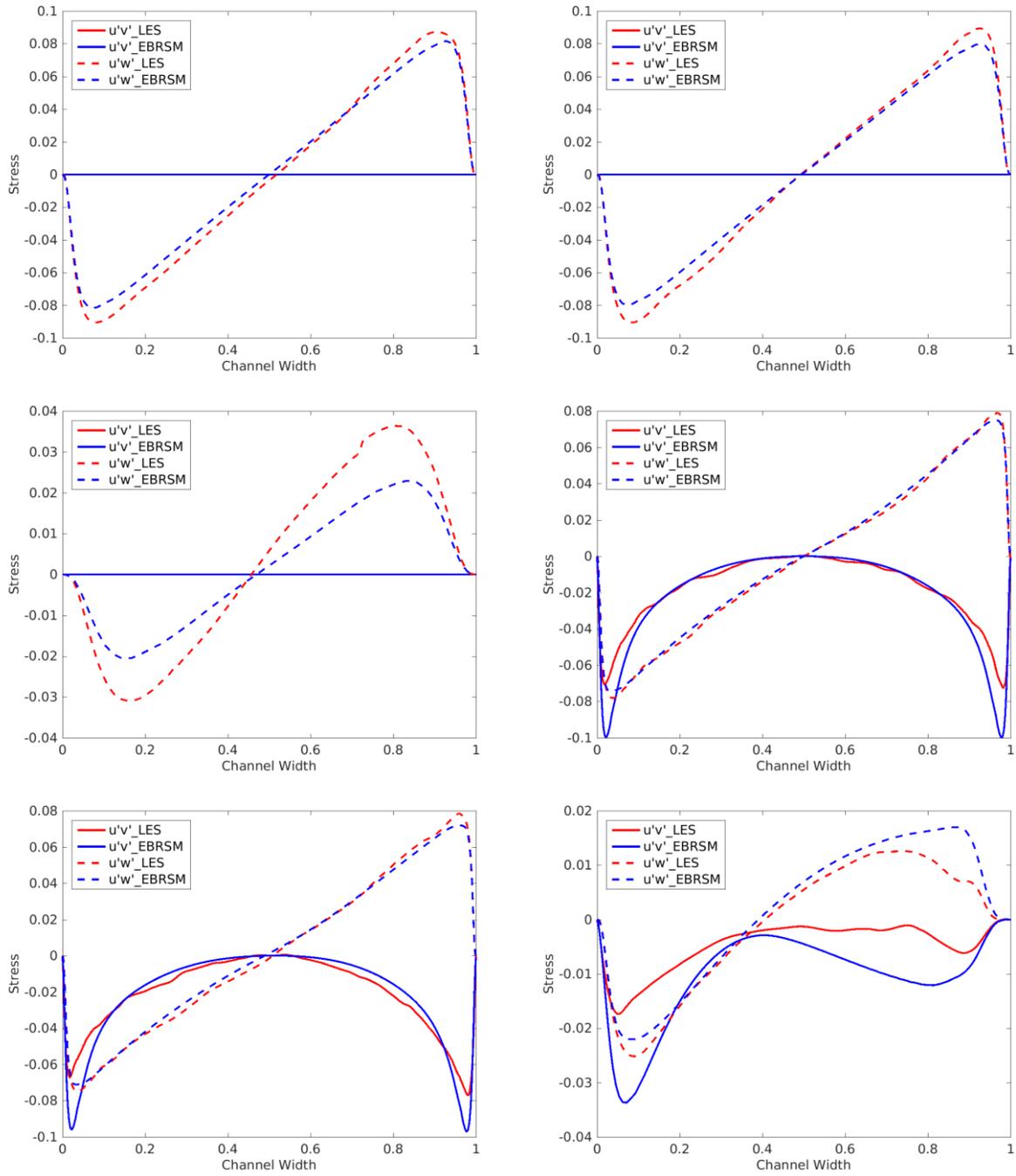

**Figure 7. Reynolds shear stresses for lines 1 (top left) through 6 (bottom right).**

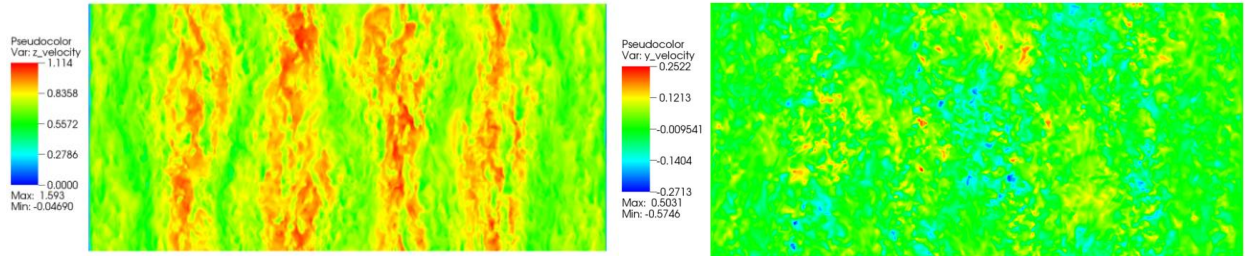

**Figure 8. Instantaneous axial velocity (left) and wall-parallel velocity (right) in the edge channel.**

Another method for quantifying the differences in flow physics and turbulence between the subchannels is *via* turbulence triangles, similar to the so-called "Lumley triangles" [16,17]. These are based on the second and third invariants of the anisotropy tensor based on the Reynolds stress tensor. The triangle bounds all realizable states of turbulence; the right corner represents pure 1-D axisymmetric turbulence, the left corner represents pure 2-D axisymmetric turbulence, and the bottom corner represents isotropic turbulence.

The half channel width data for each of the six lines are provided in Figure 9. The wide gaps in lines 1, 2, and 4 to 6 display behavior largely similar to that seen in canonical channel flow [16,18]: at low y+, the turbulence is two-dimensional on the upper triangle curve, as seen in the LES results. As wall distance increases, the turbulence becomes predominantly one-component.

In the narrow gaps, an interesting phenomenon is noted. While the center-most subchannel resembles standard channel flow, moving outward the channels display increasing shifts toward two-component turbulence in the bulk, with the edge subchannel showing very little one-component turbulence. This is a stark change in the local physics of the flow, which EBRSM is not able to capture. This is also consistent with the presence of the gap vortex street, as observed in Merzari and Ninokata [19]. This represents a major challenge for turbulence modeling, and motivates in part the forthcoming DNS.

### 4.2. Thimble Configuration

The lessons learned from the standard 5x5 case have also been applied to the thimble configuration. Approaches for this case are similar to those in the 5x5 case albeit with some adjusted meshing in the thimble region. Results shown here are preliminary but demonstrate the significant impact of a relatively small geometry change. They also display a need for DNS data to be better able to decipher the complex flow physics.

Instantaneous axial velocity plots are provided in Figure 10 for well-resolved LES. These demonstrate the blockage effect of the thimble and its substantial impact on neighboring subchannels. These narrower gaps may also be prone to instability and flow pulsation behavior as was seen in the 5x5 case. This is suggested by the large, intermittent structures seen in the central subchannels.

Figure 11 provides comparisons of Line Integral Convolution (LIC) for the 5x5 case and the thimble case for the EBRSM model. The LIC plots show the secondary flows in the plane, with the color representing the axial velocity magnitude. It is seen that in the 5x5 case, each pin has its own set of associated secondary flows. The edge channels have a separate vortex pair at the wall but in general the secondary flows are "connected" to a given pin and bounded by the unit cell. In contrast, the thimble case shows different behavior near the center. The thimble appears to have its own set of secondary flows, but the flows at the corner subchannels overlap between multiple pins. This provides some evidence of a change in the physics that, along with the potential gap vortex street in this area, may prove challenging to model. LES and DNS should be able to better demonstrate the flow physics in this area, and allow for better quantification of the EBRSM accuracy for this configuration.

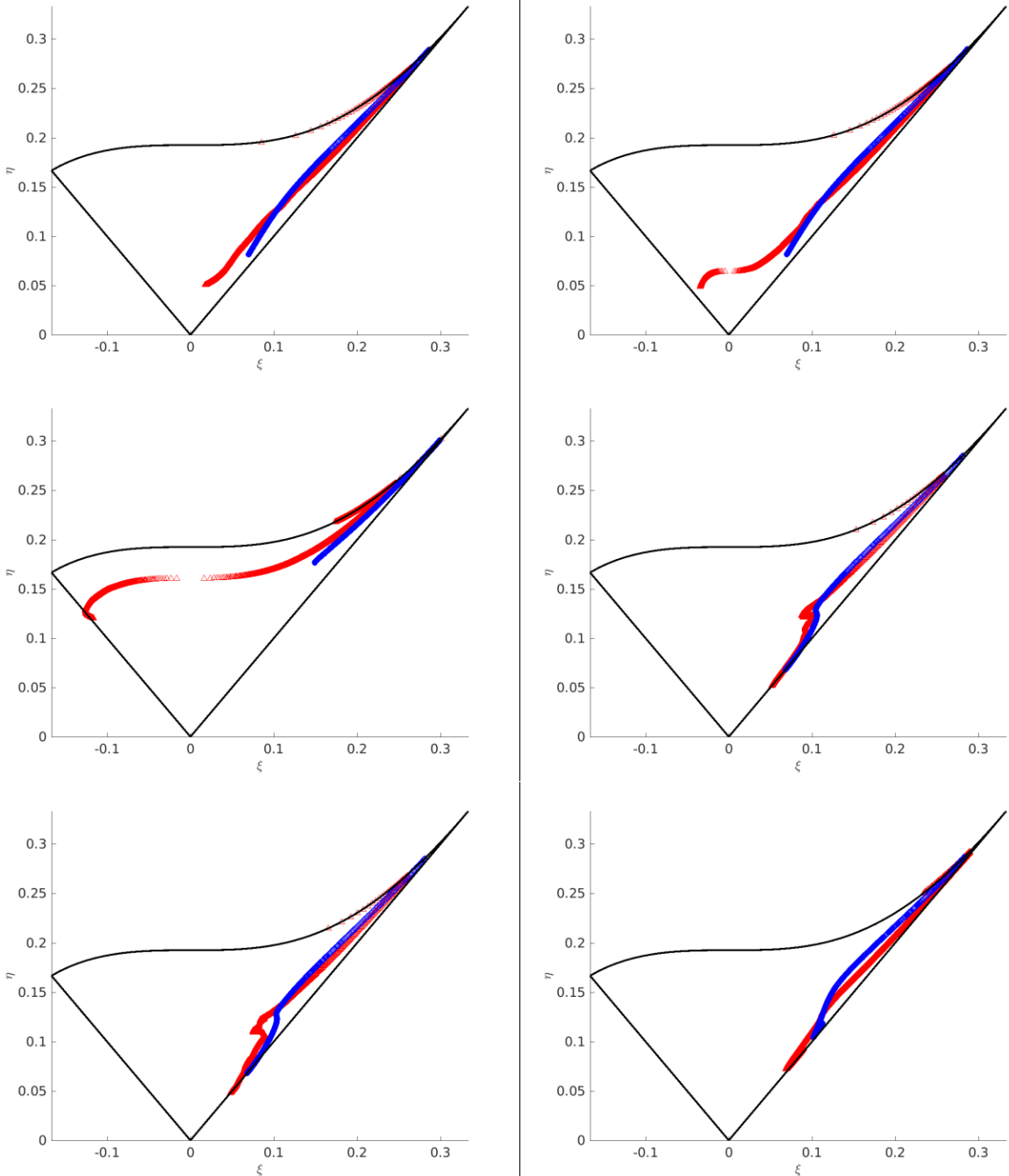

**Figure 9. Turbulence anisotropy triangles for lines 1 (top left) through 6 (bottom right). LES is red triangles, EBRSM is blue circles.**

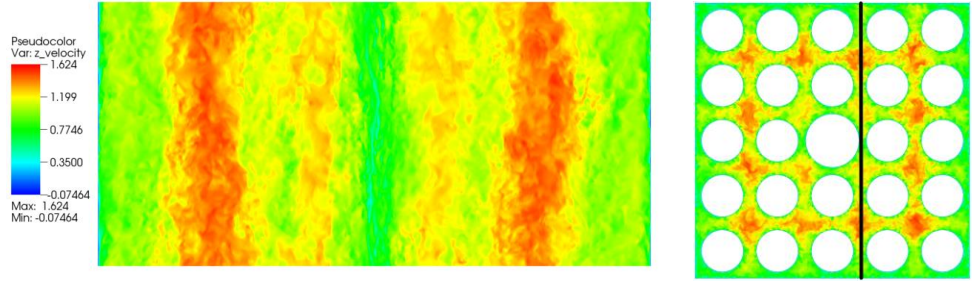

**Figure 10. Instantaneous axial velocity for the thimble case (plane on left corresponds to black line on right).**

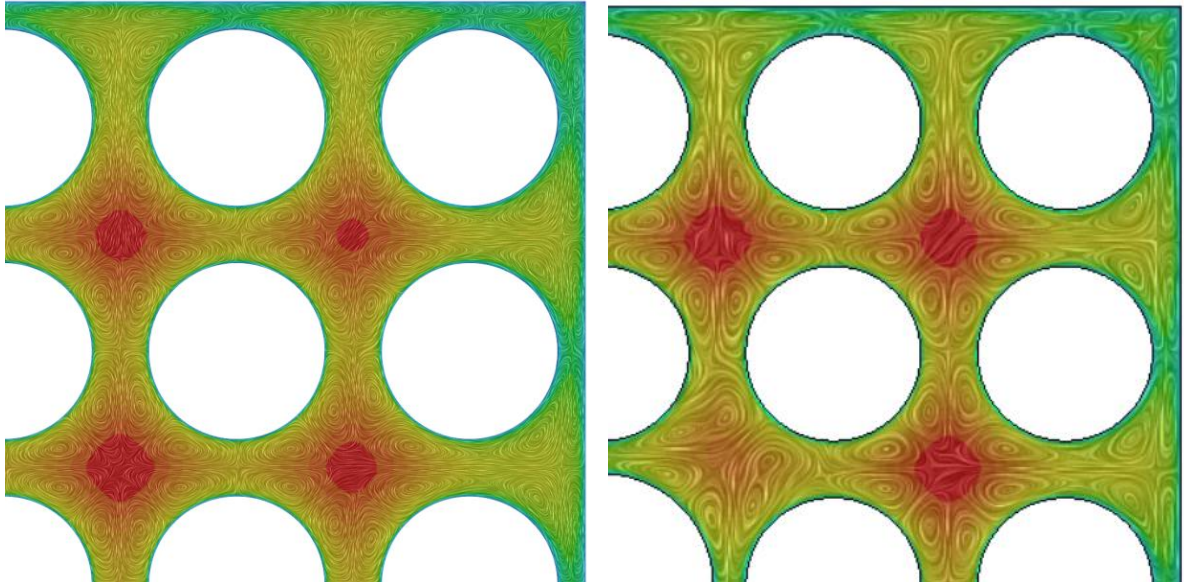

**Figure 11. Line integral convolution showing secondary flows from EBRSM for the standard case (left) and the thimble case (right).**

5.  **CONCLUSIONS**

Results from the square 5x5 and thimble cases show the general complexity of flow phenomena in rod bundles associated with the non-uniformity of the cross section. Even a very advanced RANS model such as EBRSM was shown to have difficulty predicting certain aspects of the flow, notably in the smaller gaps and edge subchannels. We observed evidence of a localized gap vortex street in the edge subchannels and a likely related shift in the anisotropy behavior. In particular in the narrow gaps of the edge channels, turbulence is two-component in nature.

We expect that the DNS for these geometries will be instrumental in gaining further insight into the physical processes associated with this phenomena. Moreover, the DNS data can be of great use for validation development and validation of various numerical models. The simulations performed here have enumerated the resolution requirements for performing DNS, as well as identified areas of particular concern. Notably, given the potential gap vortex street in the outer subchannels (and perhaps in the thimble vicinity), the domain may need to be lengthened in order to capture all relevant scales associated with this behavior.

**ACKNOWLEDGMENTS**

This research used resources of the Argonne Leadership Computing Facility, which is a DOE Office of Science User Facility supported under Contract DE-AC02-06CH11357. We also gratefully acknowledge the computing resources provided on Bebop, a high-performance computing cluster operated by the Laboratory Computing Resource Center at Argonne National Laboratory.